\documentclass[12pt]{article}

\usepackage{appendix}

\usepackage{color}
\usepackage{amssymb}
\usepackage{amsmath}
\usepackage{graphicx}
\usepackage{ifpdf}

\newcommand{\prt}{\partial_t}
\newcommand{\prx}{\partial_x}

\newcommand{\olx}{\overline{x}}

\newcommand {\MD}{M_d}

\newcommand{\vel}[0]{v}
\newcommand{\Ev}[0]{\tilde{v}}
\newcommand{\Erho}[0]{\tilde{\rho}}
\newcommand{\Ephi}[0]{\tilde{\phi}}
\newcommand{\pert}[1]{\delta{#1}}  	

\newcommand{\lx}[1]{{\mathrm{#1}}}  	
\newcommand{\pINT}[0]{\phi_{\lx{I}}}
\newcommand{\pEXT}[0]{\phi_{\lx{E}}}
\newcommand{\PEEXT}[0]{\mathcal{P}_\lx{E}}
\newcommand{\PEINT}[0]{\mathcal{P}_\lx{I}}
\newcommand{\eqn}[1]{(\ref{#1})}  	
  	
\newcommand{\Dummy}[0]{\Phi}

\date{}
\title{Stability and energetics of Bursian diodes.}
\author{M.S. Rosin,
H. Sun, Department of Math, UCLA }
\begin{document}

\maketitle

\begin{abstract}
We present an analysis of the stability, energy and torque properties of a model Bursian diode in a one dimensional Eulerian framework using the cold Euler-Poisson fluid equations. In regions of parameter space where there are two sets of equilibrium solutions for the same boundary conditions, one solution is found to be stable and the other unstable to linear perturbations. Following the linearly unstable solutions into the non-linear regime, we find they relax   to the stable equilibrium. A description of this process in terms of kinetic, potential and boundary-flux energies is given, and the relation to a Hamiltonian formulation is commented upon. A non-local torque integral theorem, relating the prescribed boundary data to the average current in the domain, is also provided. These results should prove useful for understanding Bursian diodes in general, as well as for control applications and benchmarking numerical codes.
\end{abstract}
\section{Introduction}

In its simplest form, a diode consists of two conducting electrodes with a relative electric potential bias $|\phi_1|$, and a distribution of moving charge carriers. Fundamentally, the transport of these charge carriers is constrained, self-consistently, by non-linear space charge effects. For example, in the case of a steady un-neutralized electron flow in one dimension (a ÔBursian diodeÕ), the charge flux cannot exceed the analytically derivable `Child-Langmuir limit' that depends only on $|\phi_1|$, the size of the domain and the velocity of the incoming electrons    \cite{child1911discharge, langmuir1913effect, Jaffe1944}.  Mechanistically, if the electron flux exceeds the limiting value, there is a charge build-up -- a virtual cathode -- and an associated electric field that resists the passage of additional electrons. 

Understanding and controlling the onset of this virtual cathode, as well as other, nearby, physically and numerically accessible states, their stability properties, and the energy demands of maintaining a diode flow, has applications in a wide range of settings that are well reviewed by Ender et al.  \cite{ender2000collective}. Some examples include inertial-electrostatic confinement \cite{carr2010dependence}; pinch reflex diodes for intense ion beam generation \cite{hinshelwood2011ion}; vircators \cite{sullivan1987virtual}; reflex triodes for microwave generation \cite{sharma2011development};  photoinjectors \cite{coutsias1983space,valfells2002effects}, and producing GHz to THz electromagnetic radiation \cite{akimov2001true,pedersen2010space}. 

Historically, much of the illuminating analysis has come from simulations, especially in complex geometries and for kinetic systems. 
To ensure the fidelity of future codes, a 
good understanding of the basic physics and a suite of test cases for benchmarking is desirable. Furthermore, in the fluid limit, diodes are readily analyzable, energetically open system, as the entering and exiting particles carry with them kinetic and potential energy.  They therefore constitute a good practical example from which to extend the Hamiltonian description of a fluid beyond energetically closed systems \cite{salmon1988hamiltonian, morrison1998hamiltonian}.  To these ends, this paper investigates the linear and non-linear dynamics, and the time dependent energy evolution of 
the two-equilibria region of parameter space supported by the simple Bursian diode above.

The plan of this paper is as follows. In section \ref{sec:equil} we introduce our equations and review what is known about their time-independent, i.e. equilibrium, solutions. We focus on  
regions of parameter space that supports two distinct equilibria. In section  \ref{sec:linstab} we present a new perspective  on their linear stability, showing one to be stable and the other unstable. In section \ref{sec:nonlinstab} we continue our investigation by following the linear instability into the non-linear regime, and discuss the associated system energy and torque, and their role as diagnostics.  In section \ref{sec:conc} we conclude and discuss some potential applications for our results. 

\section{Equilibrium solutions}\label{sec:equil}

A standard model to describe charge carriers in an electric potential is the cold 1D hyperbolic-elliptic Euler-Poisson system given by
\begin{eqnarray}
\partial_t \rho + \partial_x (\rho \vel) &=& 0,\label{cont} \\
\partial_t \vel +  \vel \partial_x \vel  &=& \partial_x \phi,\label{mom}\\
\partial_{xx} \phi &=& \rho \label{poisson},
\end{eqnarray}
where $x,t, \rho, \vel, \phi$ are the scaled position, time, density, velocity and potential for an electron fluid and the time independent Dirichlet boundary conditions are
\begin{align}  \label{BC1}
 \rho(x=0) = \rho_0, \quad \vel(x=0) = \vel_0, \quad \phi(x = 0) = 0, \quad \phi(x=d) = \phi_1.
 \end{align} We normalize using
\begin{eqnarray}
(x,t,\vel,\phi,\rho) = (x'/L, t'/T, \vel'/(L/T), \phi'/\varphi, \rho'/R) \nonumber \\
\varphi = (m_e/q_e) (L/T)^2 \quad \quad R=(\varepsilon_0/q_e)(\Phi /L^{2}) . \nonumber 
\end{eqnarray}
The primed variables are unscaled, $L,T$ are characteristic length and time scales,  $ m_e$ is the electron mass, $q_e$ the  fundamental charge (positive),  $\varepsilon_0$ is vacuum permittivity, and the electric field is $ -\partial_x \phi$. 

 In the steady state (\ref{cont}) and (\ref{mom}) constrain the current  $\rho \vel$ and the energy density of a fluid element, kinetic plus potential $\vel^2/2 -  \phi$, to be constant across the domain (the minus is because electrons are negatively charged). This implies that for given boundary conditions i.e. (\ref{BC1}), the two unspecified fields at the outgoing boundary $\rho(d), \vel(d)$ are uniquely determined.  The motion of the fluid can be understood energetically  in terms of Hamilton's principle, the principle of least action, from which (\ref{mom}) can be derived.  The gain (loss) in the kinetic energy of a fluid element as it crosses the domain equals its loss (gain) in potential energy as work is done on (against) it by the electric field (that accelerates electrons from the emitting cathode to the collecting anode, in the case of a monotonically increasing potential).

\begin{figure}
\centering
 \includegraphics[width =0.48\textwidth]{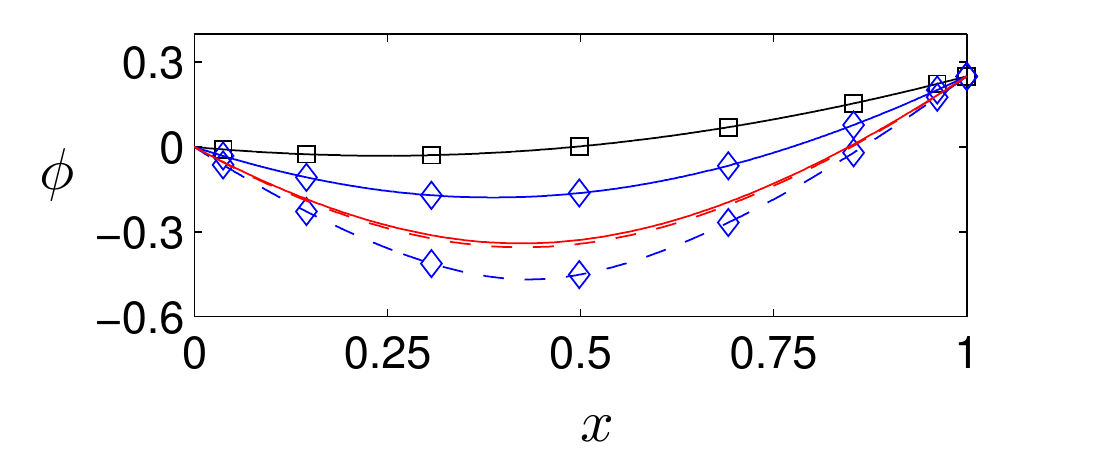}
\caption{  Potential profiles $\phi$ associated with branch I (solid lines) and branch II (dashed lines) equilibria for  $(d, \vel_0, \phi_0, \phi_1) =(1.0,1.0,0,0.25)$, and $\rho_0=2.45$ (red, no symbols), $\rho_0=2.00$ (blue, diamonds), and $\rho_0=1.00$ (black, squares) respectively. }
\label{fig:two_solutions}
\end{figure}

It is known that there are two kinds of equilibrium solutions to these equations \cite{Jaffe1944}, and we review them here.  For $\phi_1 >0$, their implicit expressions are given by:

\begin{align} 
\left(\Dummy-2\alpha\right)\sqrt{\Dummy+\alpha}=
&\frac{3}{4}\frac{\sqrt{8\rho_0}}{v_0}x+(1-2\alpha)\sqrt{1+\alpha},
&0 < d \sqrt{\frac{8\rho_0}{v_0^2}} \leq \xi_2
\label{firstkind},\\
\left(\Dummy-2\alpha\right)\sqrt{\Dummy+\alpha}=
&\left| \frac{3}{4}\frac{\sqrt{8\rho_0}}{v_0}x-(1-2\alpha)\sqrt{1+\alpha}\right|, 
&\xi_1 < d \sqrt{\frac{8\rho_0}{v_0^2}} \leq \xi_3
\label{secondkind},
\end{align}

where $\Dummy = \sqrt{1 + 2\phi/v^2_0}$ and  $\Dummy_d = \sqrt{1 + 2\phi_1/v^2_0}$ are normalized potentials,  and $\xi_1 =  4/3(1+\Dummy_d^{3/2}) <\xi_2  = 4/3 (\Dummy_d + 2) (\Dummy_d -1)^{1/2}< \xi_3 = 4/3 \left( 1 + \Dummy_d\right)^{3/2}$ demarcate (non-exclusively) the boundaries between solutions monotonic in $\phi$ given by (\ref{firstkind}) and solutions with a single turning point given by (\ref{secondkind}).

To close these equations,  $\alpha$ is needed. It satisfies
\begin{equation}
 \left(\Dummy_d-2\alpha\right)\sqrt{\Dummy_d+\alpha} = \frac{3}{4}\frac{\sqrt{8\rho_0}}{v_0}d\pm(1-2\alpha)\sqrt{1+\alpha}, 
 \label{alpha}
 \end{equation}
where the positive and negative signs correspond to (\ref{firstkind}) and (\ref{secondkind}) respectively. 

Necessary and sufficient conditions for determining the number of solutions, zero, one or two, are given succinctly by 
\begin{align}
 d \sqrt{8\rho_0 /v_0^2}&>\xi_3: \quad& &\text{zero solutions},\\
d \sqrt{8\rho_0 /v_0^2}<\xi_1 \,\text{  or  }\, d\sqrt{8\rho_0/v_0^2}&=\xi_3:\quad& &\text{one solution},\\
 \xi_1\leq d \sqrt{8\rho_0/v_0^2}&<\xi_3: \quad&  &\text{two solutions}.\label{two_solutions}
 \end{align}

\begin{figure}[tbh]
\centering
 \includegraphics[width=0.48\textwidth]{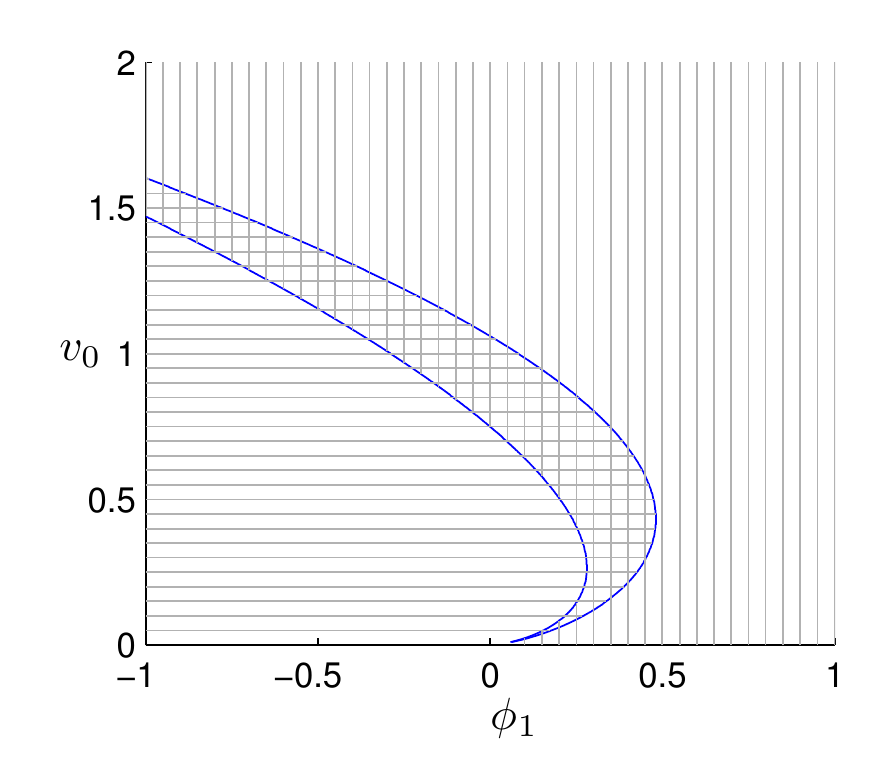}
 \includegraphics[width=0.48\textwidth]{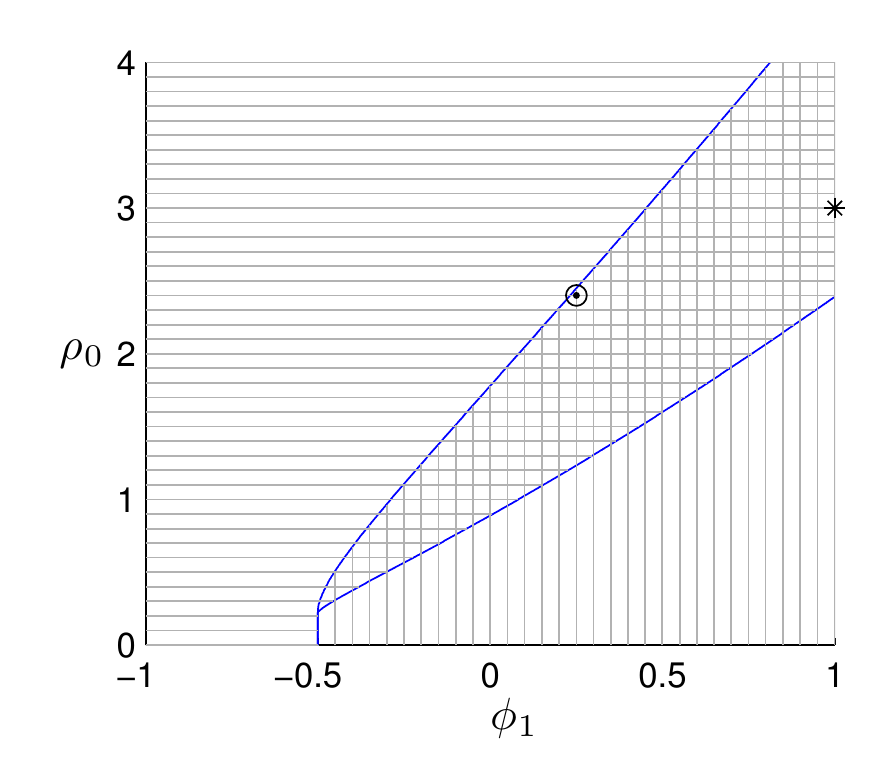}

\caption{The number of equilibrium solutions to (\ref{cont})-(\ref{poisson}): Zero - horizontal lines, one - vertical lines, two - crossing lines, for varying $\rho_0, \vel_0, \phi_1$. Left: $(d,\rho_0)= (1,1)$ 
Right: $(d,\vel_0)=(1,1)$. The zero and two solution boundary - $ d \sqrt{8\rho_0/v_0^2}=\xi_3 $ - corresponds to the space charge limiting current derived by Child and Langmuir for $v_0=0$, Jaff\'{e} for $v_0>0$, and recently, for time varying solutions, by Caflisch and Rosin  \cite{child1911discharge, langmuir1913effect, Jaffe1944, Caflisch12}.  The circled dot  corresponds to the parameters used in Fig. \ref{fig:eig_data} and the star to those used in Figs. \ref{LinEv} and \ref{EvolEnergy}\color{black}}
\label{fig:regions}
\end{figure}

The number of accessible solutions is a function of  $d$, $v_0$, $\rho_0$, and $\phi_1$, Figs. \ref{fig:two_solutions} and \ref{fig:regions}. For example, for $(d, \vel_0, \phi_1 )= (1, 1, 0.25)$, there are no steady state solution for $\rho_0>2.5$, two when $2.5>\rho_0>1.2$, and one when $\rho_0<1.2$.  We denote the solution with larger $\phi$ as branch I, the other as  branch II. In the literature, these are known as the C-branch and C-overlap branch respectively \cite{fay1938theory}.  

It is the stability, dynamics and energy of perturbations to the equilibria in the region of parameter space given by (\ref{two_solutions}), that are the focus of this paper. While these have been investigated before in a Lagrangian framework, our approach in an Eulerian framework is new, and has several advantages. Specifically,  it allows for a direct interpretation of solutions that are functions of $x$ and $t$, rather than Lagrangian coordinates; the discrete nature of the linear eigenmodes are a natural product of the formulation; and the description is robust to changes that would not allow for a Lagrangian analysis. 

 In accordance with earlier studies, we find that the C-overlap branch is unstable to linear perturbations, and we follow these into non-linear regime \cite{Jaffe1944, lomax1961unstable, kolinsky1997arbitrary}.

\section{Linear stability analysis}\label{sec:linstab}

We wish to determine the linear stability properties of branch I and branch II equilibrium solutions to (\ref{cont})-(\ref{poisson}) when (\ref{two_solutions}) holds. To do so, we use (\ref{firstkind})-(\ref{alpha}) subject to (\ref{BC1}) to construct equilibria  $\Erho(x)$, $\Ev(x)$, $\Ephi(x)$, and to these we add small perturbations  $( \pert{\rho}, \pert{v}, \pert{\phi} )=(\pert{\rho}(x), \pert{v}(x), \pert{\phi}(x))e^{\lambda t}$ that obey  $(\pert{\rho},\pert{v},\pert{\phi})=(0,0,0)$ at $x=0$ and $\pert{\phi}=0$ at $x=d$.

\begin{figure}
\centering
 \includegraphics[width=0.48\textwidth]{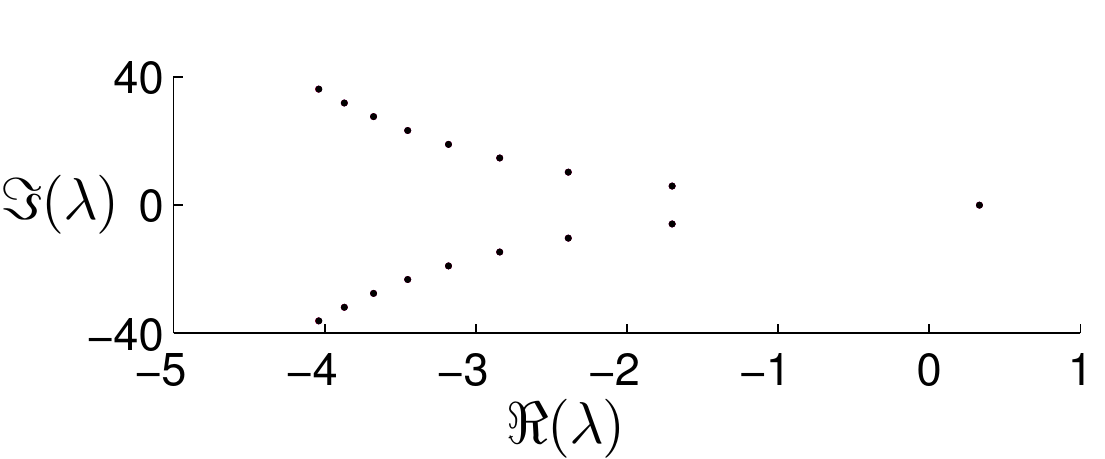}
\caption{Left: Eigenvalues associated with perturbations to branch II equilibria with parameters  $(d, \rho_0, v_0,  \phi_1)=(1,2.4, 1, 0.25)$. Only a single positive eigenvalue, the first one, exists, corresponding to an unstable, purely growing mode. The remaining eigenvalues are in complex conjugate pairs with $\Re(\lambda) <0$, corresponding to damped, traveling waves.  Results are calculated using Chebyshev spectral methods with $N=100, 200, 400 $ modes corresponding to black circles, triangles, and diamonds respectively - which overlap completely. 
 \label{fig:eig_data}}
\end{figure}

\begin{figure}
\centering
 \includegraphics[width=0.48\textwidth]{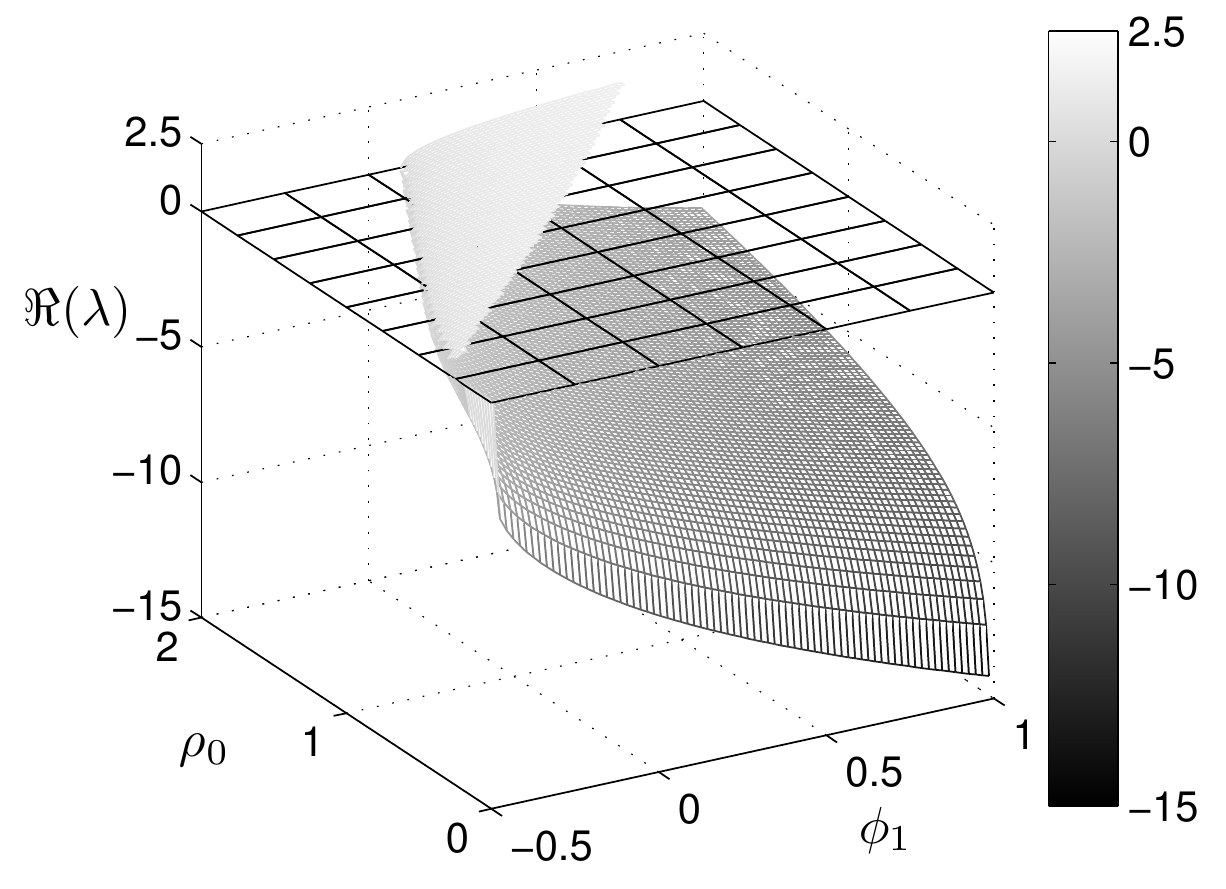}\caption{The most positive eigenvalues of (\ref{eig})  i.e. $\lambda$, associated with perturbations to branch II ($\Re(\lambda) >0$, unstable) and branch I ($\Re(\lambda) < 0$, stable) solutions for the parameters  $(v_0, d)=(1, 1)$, and varied $\rho_0$ and $\phi_1$ -- see Fig. \ref{fig:regions}. }
\label{fig:eig}
\end{figure}

Linearizing, we obtain
\begin{eqnarray}
 \lambda\pert{\rho}+\partial_x(\Erho\pert{v})+\partial_x(\Ev\pert{\rho})&=&0,\label{pert_cont}\\
 \lambda\pert{v}+\partial_x(\Ev\pert{v})&=&\partial_x\pert{\phi},\label{pert_mom}\\
 \partial_{xx}\pert{\phi}&=&\pert{\rho}. \label{pert_poisson}
\end{eqnarray}
This system can be written as
\begin{eqnarray}
 \lambda\left(\begin{array}{c}\pert{\rho}\\ \pert{v}\end{array}\right)=A\cdot\left(\begin{array}{c}\pert{\rho}\\ \pert{v}\end{array}\right), \label{eig}
\end{eqnarray}
where 
\begin{eqnarray}
 A:=\left(\begin{array}{cc}
 -\partial_x\Ev-\Ev\partial_x & -\partial_x\Erho-\Erho\partial_x \\
 \partial_x(\partial_{xx})^{-1} & -\partial_x\Ev-\Ev\partial_x
\end{array}
\right).\label{eig_sys}
\end{eqnarray}
The eigenvalues of (\ref{eig}), determine the linear stability of the system, $\Re (\lambda) > 0$ describes unstable modes, and $\Re (\lambda) <0$, stable modes. 
To compute $\lambda$, we discretize the operator matrix (\ref{eig_sys}) using three methods:  a uniform grid with an upwind scheme; a uniform grid with a centered difference scheme; and a Chebyshev grid with an associated polynomial interpolation \cite{Lloyd:2000}.  The discrete spectrum of eigenvalue-eigenvector solutions - a discreteness not generally emphasized in the dispersion relations arising from Lagrangian analyses e.g. \cite{lomax1961unstable, 
kolinsky1997arbitrary}. - are obtained numerically and shown in Figures \ref{fig:eig_data}, \ref{fig:eig}
 and  \ref{LinEv}. All three schemes converge to the same result. 

Conducting a parameter scan, for branch II we find $\lambda > 0 \in \Re$ for the first eigenvalue, the one with a single zero in the corresponding eigenfuctions. For the remaining eigenvalues in branch II, and all of branch I, $\Re (\lambda) <0$. The system supports a single unstable mode. For example, for $(d, \rho_0, v_0, \phi_1) =  (1, 1.5, 1, 0.2)$, the most positive eigenvalues from branch II and I are  $1.1$ and $-2.1$ respectively - one mode is  unstable, and the other stable.  As the two equilibrium solutions merge at  $ d \sqrt{8\rho_0/v_0^2}=\xi_3 $, the unstable eigenvalue of branch II obeys $\Re(\lambda) \to 0$. Approaching the other boundary of the two solution region  $ d \sqrt{8\rho_0/v_0^2}=\xi_1$, the full-width, half-maximum of the corresponding eigenmode $\to 0$. It remains to be seen whether this singular mode bears any fundamental relation to the singularity that forms in the case that the current exceeds the Child-Langmuir limit \cite{coutsias1988caustics, Caflisch12}.

\section{Nonlinear dynamics}\label{sec:nonlinstab}

For small time, coupling between infinitesimal amplitude perturbations, and their feedback on the equilibrium solutions, is negligible. However, because $\lambda > 0$ for one mode, that mode grows exponentially and the perturbations quickly reach non-linear amplitudes. In this case,  the methods and results of section \ref{sec:linstab} are no longer applicable. In the non-linear regime, the most general method for  solving (\ref{cont})-(\ref{poisson}) is numerical integration; although the method of characteristics can also be used to obtain complete analytic solutions in a Lagrangian framework \cite{kolinsky1997arbitrary, ender2000collective}. The method used here, an Eulerian approach, has the advantage that it is naturally formulated as a two point Dirichlet boundary value problem for $\phi$, which can easily be realized experimentally. The alternative Lagrangian approach is more naturally formulated as a Cauchy problem including $\partial_x \phi$, which is harder to realize experimentally, and from which the corresponding Dirichlet conditions are non-trivial to obtain  \cite{Caflisch12}.

\begin{figure} [tbh]
 \centering
 \includegraphics[width=0.48\textwidth]{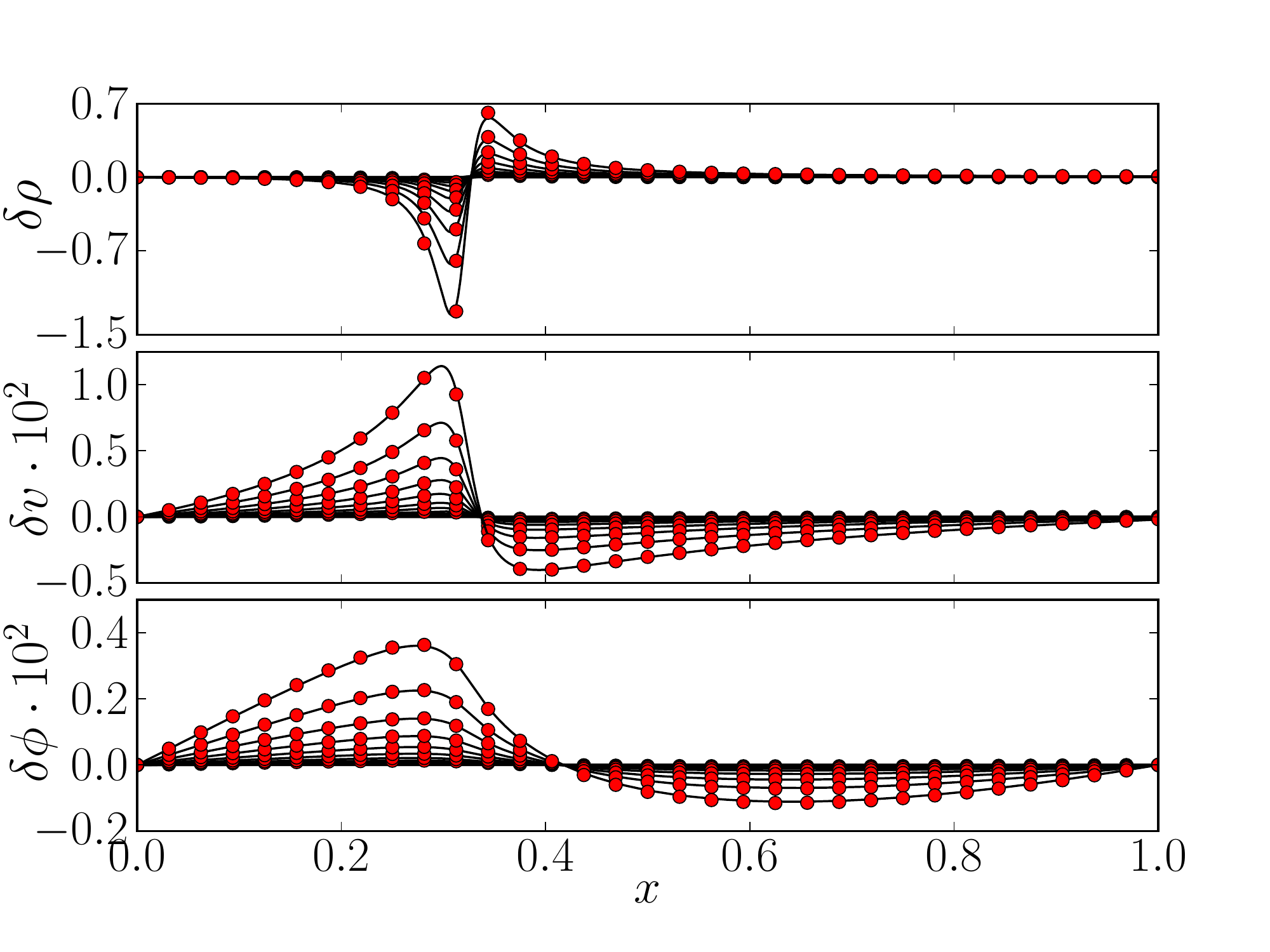}%
 \includegraphics[width=0.48\textwidth]{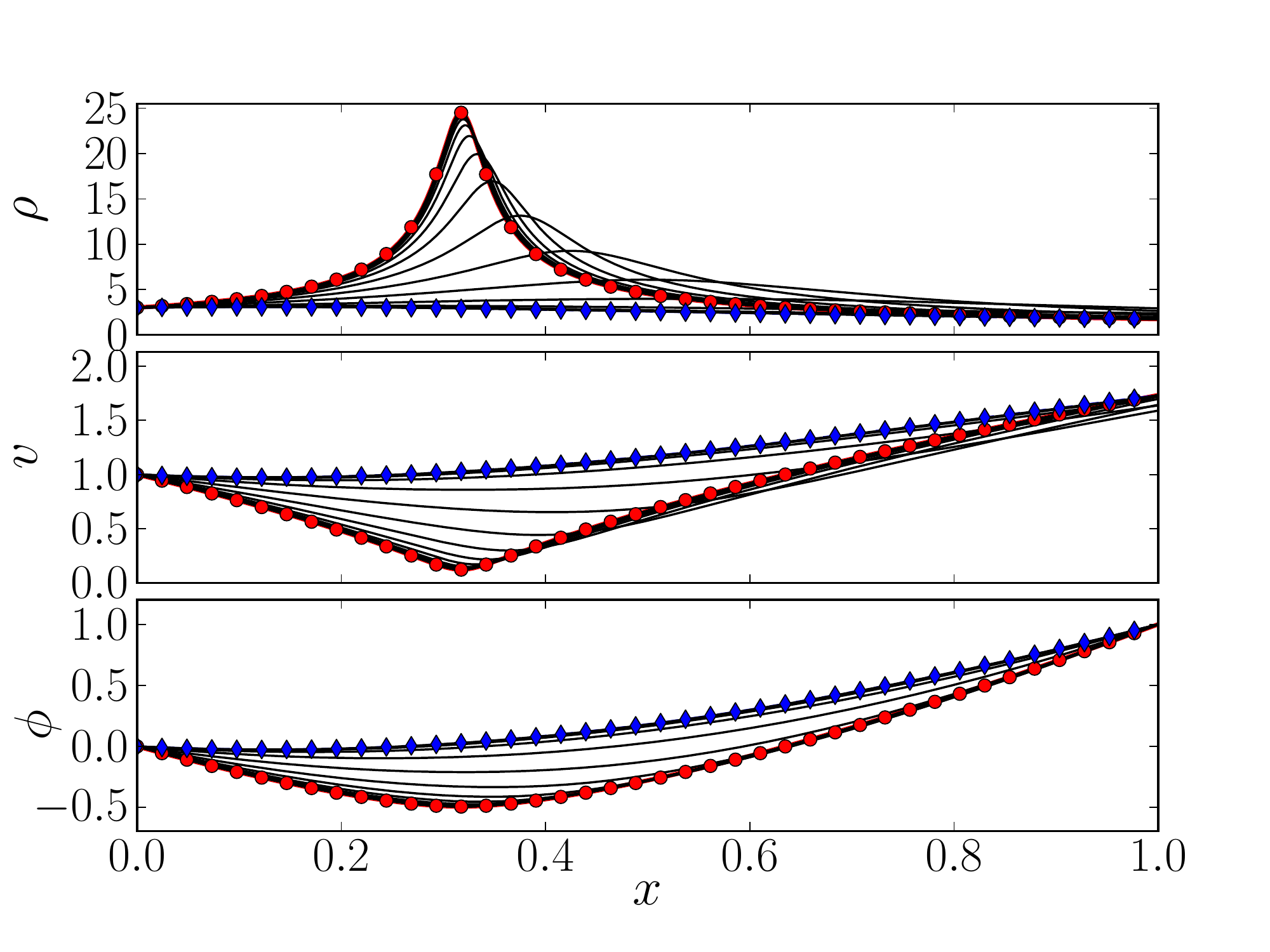}%
  \caption {Left hand side: Evolution of perturbed field quantities associated with the unstable branch II equilibrium for  $t \leq 5.6$ and steady boundary conditions  $( d,    \rho_0, v_0,  \phi_1) = ( 1 ,  3, 1, 1)$. Snapshots are every $t =0.35$ starting from the $\pert{\rho},\pert{v}, \pert{\phi} = 0$ initial conditions. The unstable linear eigenmodes (red squares) with growth rate $\exp(\lambda t)$ - section \ref{sec:linstab} -  match the fully non-linear solutions for small time. Right hand side:  Evolution of full solutions to (\ref{cont})-(\ref{poisson}) starting at the same branch II equilibrium, for  $t \leq 11.5$. Snapshots are every $t = 0.5$, and the initial state (red circles) is given by (\ref{firstkind}) - (\ref{alpha}). The final state (blue diamonds) is the same stable branch I equilibrium derived from the same set of equations and boundary conditions as the initial state. }
     \label{LinEv}
\end{figure}

We favor the numerical approach. We employ MacCormack's method to integrate the hyperbolic equations (\ref{cont})-(\ref{mom}), and solve the elliptic Poisson equation (\ref{poisson}) at each time step using a finite-difference description and inverting a tridiagonal matrix. Our simulations are initialized with  unstable equilibrium solutions from branch II and numerical noise provides broadband perturbations which are constrained to obey (\ref{BC1}). The solutions to our perturbed system are well matched by our linear results for small time, and in the final state  the solutions have relaxed to the stable branch I equilibrium solutions with the same boundary conditions as the initial, unstable equilibrium, Fig. \ref{LinEv}. 

Physical insight  and a set of numerical benchmarks for the system, can be obtained by considering both the energetics of the 
system and its global torque. In the next section, we examine each in turn, and derive a set of integral equations that describe the system's spatially averaged properties 
and their interaction with the boundaries. 

These type of equations are both less computationally demanding to solve (which is unimportant here, but may matter in higher dimensions or kinetic models), and do not require knowledge of the fundamental unaveraged solutions.  Furthermore, being able to 
relate \emph{prescribed} boundary value data to \emph{derived} domain data offers a new avenue for both control, and experimental measurement 
of spatially-distributed system properties. 

\subsection{Energetics}

Even at the model equation level considered here, energy insights may be important for industrial purposes \cite{sullivan1987virtual}.  In this section, we examine the evolution and balance of the standard energy integrals. We leave further detailed discussion to a forthcoming paper in which we present a tailored Bursian diode-battery model  \cite{Rosin12CLEnergy}.

We start by multiplying (\ref{mom}) by $\vel$ and combining it with (\ref{cont}) to yield an 
evolution equation for the kinetic energy $\mathcal{K} = \rho \vel^2/2$ balance of the system
\begin{align} \label{KE}
\partial_t \mathcal{K} + \partial_x \left( v \mathcal{K} \right) = \rho v \partial_x \phi,
\end{align}
where $\rho v \partial_x \phi$ is the negative Joule heating term. 
Integrating over $x$, the total kinetic energy in the system is given by
\begin{align} \label{KEint}
\partial_t \overline{\mathcal{K}}   =    v_0 \mathcal{K}_0 -v_d \mathcal{K}_d + \overline{\rho v \partial_x \phi} 
\end{align}
where over-bars denotes spatially integrated quantities $\int^d_0 dx$ and subscripts $0, d$ indicate that the associated quantity is to be evaluated at $x = 0,d$ respectively. There are two contributions to the total kinetic energy: the difference in the boundary fluxes of  kinetic energy, and the work done on the fluid by the electric field.

To describe the total energy balance in the system, it helps to decompose $\phi  = \pEXT + \pINT$ into  external and internal components, and these satisfy Laplace's and Poisson's equations respectively:
\begin{align}
\partial_{xx} \pEXT &= 0,  \lx{\;with\;}\pEXT(0) = 0, \;  \pEXT(d) = \phi_1  \label{intpoisson}\\
\partial_{xx} \pINT &= \rho,  \lx{\;with\;}\pINT(0) = 0, \; \pINT(d)  = 0.\label{extpoisson}
\end{align} 
The solution to (\ref{intpoisson}) is simply $\pEXT = (\phi_1/d) x$, and the Green's function for $\pINT$ is
\begin{align}
\pINT(x) = \frac{1}{2} \int^d_0 dx' \rho(x') \left( |x-x'| - 2\frac{ x x'}{d} - x - x' \right).\label{Greens}
\end{align}

 \begin{figure}
 \centering
  \includegraphics[width=0.48\textwidth]{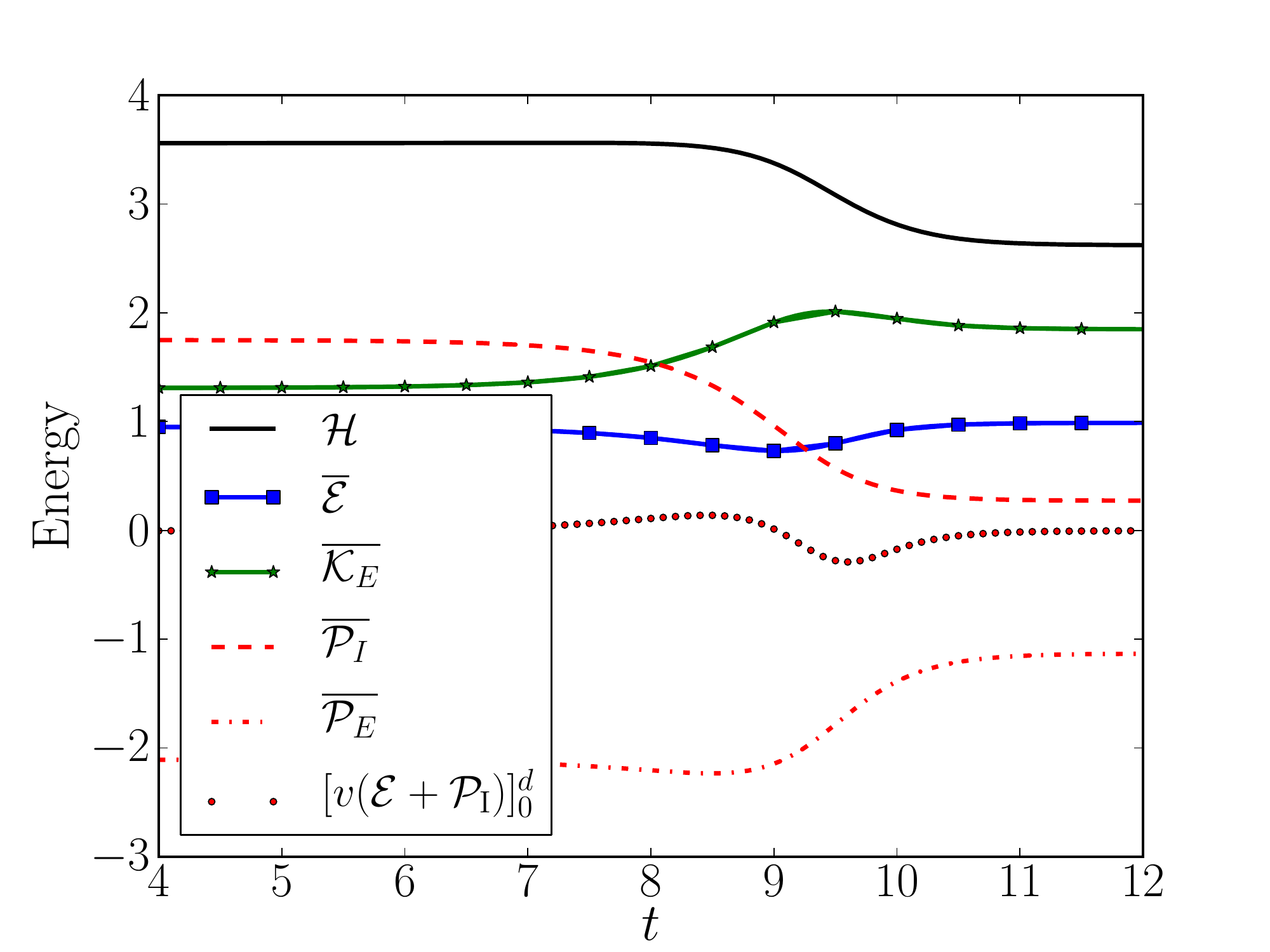}%
  \caption {Evolution of the total (spatially integrated) energy of a Bursian diode system from an initial unstable equilibrium  solution (branch I) to a final stable equilibrium solution (branch II) for the same set of boundary conditions as in Fig. \ref{LinEv}. Whilst the final total energy state $\overline{\mathcal{E}}$ is slightly greater than the initial state, the Hamiltonian $\mathcal{H}$ is a strictly decreasing function of time. The system's dominant form of energy switches  to kinetic  $\overline{\mathcal{K}}$ from internal potential energy $\overline{\mathcal{P}_I}$ as time progresses.}
\label{EvolEnergy}
\end{figure}

Rewriting (\ref{KE}) in conservative form using (\ref{cont}) and (\ref{extpoisson}), we have
\begin{align}
\partial_t  \mathcal{E}  + \partial_x \left(\vel \left( \mathcal{E} + \PEINT \right) \right) = 0,
\label{totalenergy}\end{align}
where $\mathcal{E} = \mathcal{K} + \PEEXT + \PEINT$ is the combined kinetic  $\mathcal{K}$, external potential  $\PEEXT = -\rho \pEXT$  and internal potential   $\PEINT = -\rho \pINT/2$ energy of a fluid element, and  we have made use of the fact that  $\pEXT, \pINT(0)$ and $\pINT(d)$ are time independent. Physically, the factor of a half in the definition of $\PEINT$ is to avoid double counting particle interaction energies \cite{jackson1965classical}. Mathematically, it arises from the symmetry properties of  the Green's function (\ref{Greens}) under $x \Leftrightarrow x'$. 

In the absence of net boundary fluxes, the second term in (\ref{totalenergy}) vanishes upon integration. In this case, the total energy $\overline{\mathcal{E}}$ is conserved, and coincides with the fluid Hamiltonian $\mathcal{H} = \overline{\rho \vel^2/2  + (\partial_x \phi)^2/2}$, from which the equation of motion (\ref{mom}) can be derived \cite{holm1986hamiltonian, akimov2001true}.  The evolution of the various energy quantities is plotted in Fig. \ref{EvolEnergy}. 

Considerable work has been done on the non-linear stability of closed plasma and fluid systems using variational principles e.g. \cite{bernstein1958energy, holm1985non-linear, morrison87vlasovpoisson, rein2003non}. However, for open systems i.e. ones with sources, like boundary fluxes, stability proofs are difficult to construct, and we do not attempt to do so here. Nevertheless, the non-linear stability and Hamiltonian structure of such systems has been the focus of recent work, and so a theorem tailored to the problem described here may be forthcoming \cite{van2002hamiltonian,jeltsema2007pseudo, jeltsema2009lagrangian, nishidahamiltonian2012}.

\subsection{Torque and boundary conditions.}

Unlike energy, torque is not generally considered as an important property of diode systems. However, it is frequently invoked in describing stellar systems governed by  (\ref{cont}) - (\ref{poisson}), in the context of which (\ref{mom}) is known as the Jeans equation, and $\phi$ is the gravitational potential. We consider it here too and derive a simplified lower moment analogue to the astrophysical virial theorem including  boundary effects \cite{chandrasekhar1964higher}. As for the virial theorem, we find a `basic structural relation that the system 
must obey' \cite{collins1978virial}. 

To proceed, we note that, uniquely,  the 1D version of Poisson's equation (\ref{poisson}) can be directly integrated to yield 
\begin{eqnarray}
\int^x_0 dx \partial_{xx} \phi = \partial_x \phi(x) - \partial_x \phi(0)  =   \int^x_0 dx' \rho(x') = M(x) := M_x, \label{eq2}
\end{eqnarray}
which is the mass between $0$ and $x$ (which can vary with time).  It follows from \eqn{eq2} that 
\begin{eqnarray}
\phi_1 = \int^d_0 dx \left( \int^x_0 dx' \rho(x') + \partial_x \phi(0)\right)  = \MD\left( d  - \overline{x} \right) + \partial_x \phi(0) d, \label{simple}
\end{eqnarray}
where $\overline{x} \equiv \int^d_0 dx x \rho /  \int^d_0 dx \rho $ is, by definition, the center of mass, and $\MD$ is defined in (\ref{eq2}).

 Equation (\ref{simple}) has a very simple interpretation. By definition, the torque about a point $d$ is $T = F r$ where $r$ is the magnitude of the directional vector joining $d$ and the point at which $F$, the force perpendicular to this vector, acts. We consider a force acting at and proportional to the system's center of mass $\MD$, and perpendicular to $\nabla x$, say a gravitational force $F = \MD g$. In this case, we have $T = \MD g (d- \overline{x})$, and so
\begin{eqnarray}
\phi_1 -   \partial_x \phi(0) d  = (d-\olx) \MD \equiv  T, \label{torque}
\end{eqnarray}
where we have absorbed $g$ into the definition of $T$.  For time independent $\phi_1 - \partial_x \phi(0) d$, this implies that the total torque on the system is constant.  

This results in an interesting relation between the current, the rate of change of the incoming electric field $\partial_t \partial_x \phi(0)$ and exiting potential $\partial_t \phi_1$. Differentiating (\ref{torque}),
\begin{eqnarray}
\prt T =  - \prt \olx \MD + \left( d - \olx \right) \prt \MD = \partial_t \left[ \phi_1  - \partial_x \phi(0) d  \right], \label{Torque} 
\end{eqnarray}
 and, using (\ref{cont}), we have
 \begin{eqnarray}
 \prt \MD &=&  - \int^d_0 dx \, \prx (\rho u) = \rho_0 u_0 - \rho_d u_d, \label{Mass} \\
\prt \olx  \MD &=& - d \rho_d u_d + J(d) - \prt \MD \;\olx, \label{COM2}
\end{eqnarray}
where $J(d) \equiv \int^d_0 dx \, \rho u$ is the current in the domain, and (\ref{Mass}) simply states that the rate of change of mass is the flux in minus the flux out.

Combining (\ref{Torque}), (\ref{Mass}) and (\ref{COM2}) we find
\begin{eqnarray}
\rho_0 u_0 = d^{-1} \left(J(d) + \partial_t \phi_1\right) - \partial_t \partial_x \phi(0), \label{av_j2}
\end{eqnarray}
which is the main result of this section\footnote{An alternative derivation of this result due to R.Caflisch can be obtained by simply evaluating the Green's function solution of (\ref{poisson}) for $\phi$ at $x=d$ with appropriate boundary conditions - private correspondence.}.  

Equation (\ref{av_j2})  relates the average current in the domain, a derived quantity, to a set of boundary data. This, potentially, 
affords a new avenue for control. As mentioned earlier, because (\ref{cont}) - (\ref{poisson}) can be written in
characteristic form, in a mathematical sense, the appearance of the incoming electric field $-\partial_x \phi(0)$ is a more natural choice of boundary 
condition than $\phi_1$.

\section{Conclusion}\label{sec:conc}
While Bursian diodes have been well studied over the last century, the advent of large scale, multi-dimensional particle in cell codes, and fluid codes in complex geometries have the potential to offer new insights into their basic physics and to guide their design. The work provided here, whilst relatively basic in that it is one dimensional and uses a minimal set of equations, is thorough and, as such, provides a reliable set of results and integral theorems against which the results of simulations of more complicated systems can be compared. Where we have employed numerical tools we have cross checked our results using multiple methods and conducted appropriate convergence studies. 

Our results include a linear stability analysis of the unstable branch II equilibrium (C-overlap branch), and non-linear simulations of its evolution. We have found that its relaxed state is that of the stable branch I equilibrium with the same boundary conditions.  We have also provided a quantitative discussion of the role of energy and torque in diagnosing and controlling the system, and, to the best of our knowledge, our interpretation of the latter is new in the literature. 

Possible extensions to this work include constructing a non-linear stability theorem in the spirit of Bernstein et al., but including boundary fluxes \cite{bernstein1958energy}; using the results herein for benchmarking more complicated systems including investigating the stability of Child-Langmuir limited solutions; and prescribing optimizing and efficiency enhancing conditions or frameworks for diode operation \cite{griswold2010upper, Caflisch12}.

\section{Acknowledgments}
Special thanks to R. Caflisch for helpful suggestions throughout and LLNL's Visiting Scientist Program for hosting MSR. Also thanks to C. Anderson,  B. Cohen,  A. Dimits, M. Dorf, T. Heinemen,  J. Hannay, S. Lee, L. LoDestro, A. Mestel, P. Morrison,  D. Ryutov and D. Uminsky for useful conversations, and hello to Jason Isaacs. 
This work was funded by Department of Energy through Grant No. DE-FG02-05ER25710 (MSR), the Air Force Office of Scientific Research STTR program through Grant No. FA9550-09-C-0115 (HS) and NSF grant DMS-0907931 (HS).

\bibliographystyle{nar}
\newcommand{\mnras}[0]{MNRAS}\newcommand{\apj}[0]{ApJ}\newcommand{\apjs}[0]{ApJ}\newcommand{\apjl}[0]{ApJ
  Letters}\newcommand{\araa}[0]{ARA\&A}\newcommand{\aap}[0]{A\&A}

\end{document}